# New Schemes for Self-Testing RAM

Gh. Bodean, D.Bodean, A.Labunetz
Technical University of Moldova
St. cel Mare 168; MD2012 Kishinau, R. Moldova
gbodean@mail.md

**Abstract**

*This paper gives an overview of a new technique, named pseudo-ring testing (PRT). PRT can be applied for testing wide type of random access memories (RAM): bit- or word-oriented and single- or dual-port RAM's. An essential particularity of the proposed methodology is the emulation of a linear automaton over Galois field by memory own components.*

## 1. Introduction

March algorithms are commonly and widely used to test the units of random access memory (RAM). In [1] it is proposed a formal notation of the mathematical presentation of March-test algorithm. An example of such notation is the following: MarchA= $\{\updownarrow(w0); \Uparrow(r0w1); \Downarrow(r1w0)\}$, where the symbols $\{\Uparrow, \Downarrow, \updownarrow\}$ refer to traversing the address space of a memory (respectively up, down or don't care address order). The operations $w_d$ and $r_d$ refer to write or read logic value: d∈ {0, 1}.

A new self-test RAM methodology (technique), named *pseudo-ring* testing (PRT), was proposed recently [2]. In this paper a retrospectives and some new schemes of this technique for a wide type of RAM are presented. PRT schemes will be referred to *bit-oriented* (BOM) and *word-oriented memory* (WOM). The cell size $m$ of the BOM is equal to 1 and for WOM is $m>1$. Some algorithmic features for *single* and *multi-port* memories also will be presented.

## 2. PRT-scheme for BOM and WOM

The PR-testing consists of some, so-called, π-test iterations. In particular case, such iteration can be shown by the equation:

$$\pi - \text{iteration} = \updownarrow_{i=0}^{n=2} \{r_i, r_{i+1}, w_{i+2}(r_i \oplus r_{i+1})\}, \quad (1)$$

where {…} signifies a sub-iteration that performs the included operations. Index $i$ refers to the address cell of a memory array. Symbol ⊕ signs the operation *sum modulo* 2 (XOR). In this case, test iteration (1) starts with initialization of the first $k=2$ cells, where are written d. For BOM d is a logic value and for WOM d∈ $\{0,...,2^m-1\}$. In a WOM March scheme values d derive heuristically.

In the next step first and second read-operations read from two neighboring memory cells and the third write-operation writes the sum modulo 2 (XOR) of the read values to the next (third) memory array cell. Next set of operations will read second and third memory cells and writes results of its logical contains to the next, forth, memory cell, and so on. This enables an *automatic* generation of the *test data background* (TDB).

At the end of π-test iteration the final state *Fin* is compared with expected one $Fin^*$. The state $Fin^*$ can be estimated a priori, because π-test iteration has an adequate linear automaton model. This model is well known as *Linear Feedback Shift Register* (LFSR). The LFSR can be *bit-oriented*, when the stage contains one bit, and *word-oriented*, when the stage contains more than one bit.

Structure of LFSR is defined by polynomial $g(x) = \sum_{i=0}^{k} a_i x^j$, where $k$ is the number of linear automaton register stages; $a_i$ are the Galois field (GF) elements.

The PRT technique can be illustrated as a transition of a (*virtual*) linear automaton in the space of the cells memory array. Figure 1 illustrates the expected states of the BOM and WOM cells after iterative execution of sub-iterations (1). For the scheme of figure 1,b the word-oriented *virtual* LFSR is defined by polynomial $g(x)=1+2x+2x^2$ that is irreducible in the field **GF**$(2^4)$; size of word $m$ is equal to 4. In each sub-iteration (1) is performed the multiplication and *sum modulo* $p(z)$, where

$p(z) = \sum_{i=0}^{m} b_i z^i$ is an irreducible polynomial over **GF**(2), $b_i \in \{0,1\}$. For given example $p(z)=1+z+z^4$.

Multiplication over Galois field extensions is a more complex operation. It's proposed an algorithm to design the optimal scheme of multiplication by a constant in GF.

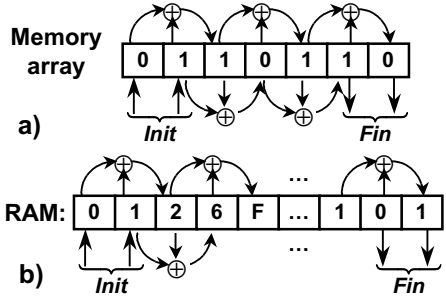

**Figure 1.** A π-test iteration: (a) for a BOM and (b) for a WOM.

Multiplier by a constant contains only XOR-gates and can be implemented inherently in the memory circuit.

If the memory array size is multiple by the period of LFSR then virtual automaton will return to the initial state at the end of the test iteration (see figure 1, b). So, the π-test quality will be estimated by comparing initial *Init* and final *Fin* states of LFSR.

For the WOM there are *intra-word* faults that can be tested by parallel application of a π-testing for BOM. In this case it is supposed that there are *m* independent bit-oriented linear automatons. For all automatons the read and write operations are executed simultaneously. To detect the intra-word faults two different π-testing can be performed: (1) with parallel or (2) with random trajectories. The trajectory is controlled by a small hardware overhead that can be programmed externally.

## 3. Pseudo-ring test quality

The quality (faults coverage) of PRT RAM can be strongly estimated by mathematical analysis of LFSR behavior in the space of a memory array. Defects in memory array are (generally) due to shorts and opens in memory cells, address decoder and read/write logic. These defects can be modeled as single and multi-cell memory faults [1]. There are three factors that influence on π-test quality and they can be used to control the π-testing: 1 – LFSR structure that is determined by generator polynomial structure; 2 – initial values for the π-test iterations; and 3 – LFSR trajectory that can be random, where address of memory cells are randomly selected, or deterministic, where address cells are selected in an increasing or a decreasing mode.

Applying Markov chain analysis it was shown that π-test iteration has a high resolution for most memory faults. Also, the combinatorial analysis of RAM π-testing has shown that it is necessary at least 3 test iteration to cover all specified faults. In [2] was proved that all single and multi-cell memory faults are detected in 3 π-test iterations with a specific TDB. This result is valid for bit-oriented as well as for word-oriented memories. The time complexity of π-test iteration for single port RAM is order $O(3n)$. This time can be reduced in the case of multi-port RAM PRT.

## 4. PRT-scheme for a multi-port RAM

Two-port (2P) memory has two independent ports. Each port has an address and control input and data input/output. The read-write operation can be performed simultaneously. To implement π-test technique for 2P memories an additional hardware overhead on RAM chip area is need: "conversion" of the existent address registers into counters and a specific XOR-logic. The ponder of the hardware overhead in comparison with the memory capacity is of an order $<2^{-20}$.

When polynomial $g(x)$ has 2 terms is recommended (see [3]) to implement the π-testing scheme, shown in figure 2. The scheme represents the evolution of 2-stages LFSR. Symbol ⊕ signs the *sum modulo p(z)*. Output edge means the *read* operation, and input edge - *write* operation. Edges on one side of the vertical axis, that pass through the cells $c_i$ ($i= 0,…, n-1$), are referred to operation on the same port. The read operations of the scheme of figure 2 are carried out simultaneously. Thus, the time complexity of a π-test iteration for the analyzed schemes is equal 2n.

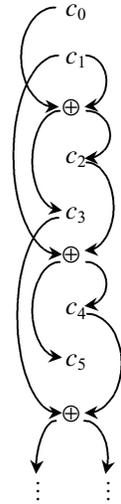

**Fig. 2. Scheme of 2P RAM PRT.**

More complex and sophisticated PRT schemes can be designed for the QuardPort DSE family. For this type of units can be designed a single or multi-LFSR schemes. Choose of the π-testing scheme depends on the fault models.

Generally, exists a high degree of mobility to control the π-test experiments. The distinct feature of the pseudo-ring testing can be declared by the paradigm: ***testing memory by its own components***. In this test schemes the RAM components, namely the memory array cells, take an active part in the test experiments.

## References

[1] A.J. van de Goor. Testing Semiconductor Memories, Theory and Practice. *ComTex Publishing*, 1998.
[2] Gh.Bodean. PRT: Pseudo-Ring Testing - a Method for Self-Testing RAM. In *Proc. Int'l Test Conf.* AQTR 2002 (THETA 13), Cluj-Napoca, Romania, 2002.
[3] Gh.Bodean. Pseudo-Ring testing scheme of word-oriented dual-port RAM. In *Proc.of the 9th International Conference on Optimization of Electrical and Electronic Equipments* (OPTIM 2004), Brasov, Romania, 2004.